\documentclass[12pt]{article}

\usepackage{amsmath,amsfonts,amssymb,latexsym,graphicx,hyphenat}

\usepackage{caption}
\usepackage{subcaption}

\def\be{\begin{equation}}
\def\ee{\end{equation}}
\def\bq{\begin{eqnarray}}
\def\eq{\end{eqnarray}}
\def\beq{\begin{eqnarray}}
\def\eeq{\end{eqnarray}}

\begin{document}
\title{\textsc{The crease flow on null hypersurfaces}}
\author{{\Large{\textsc{Spiros Cotsakis}}$^{1,2,3,}$}\thanks{\texttt{skot@aegean.gr}}\\
$^{1}$Department of Applied Mathematics and Theoretical Physics \\ University  of Cambridge \\
Wilberforce Road, Cambridge CB3 0WA, United Kingdom \\ \\
$^{2}$Clare Hall, University of Cambridge, \\
Herschel Road, Cambridge CB3 9AL, United Kingdom\\  \\
$^{3}$Institute of Gravitation and Cosmology\\  RUDN University\\
ul. Miklukho-Maklaya 6, Moscow 117198, Russia}
\date{March  2024}
\maketitle
\newpage
\begin{abstract}
\noindent The crease flow, replacing the Hamiltonian system used for the evolution of crease sets on black hole horizons, is introduced and its bifurcation properties for null hypersurfaces are discussed. We state the conditions of nondegeneracy and typicality for the crease submanifolds, and find their normal forms and versal unfoldings  (codimension 3). The allowed boundary singularities are thus prescribed by the Arnold-Kazaryan-Shcherbak theorem for 3-parameter versal families, and hence identified as swallowtails and Whitney umbrellas of  particular kinds. We further present the bifurcation diagrams describing crease evolution at the crossings of the bifurcation sets and elsewhere,  and a typical example is studied. Some remarks on the connection of these results to the crease evolution on black hole horizons are also  given.
\end{abstract}

\newpage
\section{Introduction}
In the recent paper \cite{cot23}, it was shown how to study bifurcation theory effects  associated with gravitational collapse and the formation of spacetime singularities and their topological metamorphoses that occur due to the inherent instability of the gravitational interaction. The problem of studying `crease structures' (defined as endpoints of at least two horizon generators) on the event horizon of a generic black hole  as 2-dimensional spacelike submanifolds as well as their physical importance  for the  appearance and instantaneous change  in black hole mergers and horizon nucleation or annihilation,  has been addressed in many papers using a `totally fictitious' Hamiltonian dynamical system,  starting with \cite{frie},    and further explored in  \cite{stew91,kro,gr23} and refs. therein.

In this Letter, using methods similar to those in \cite{cot23} (in small codimension up to 3), we shall further study the evolution of crease submanifolds by introducing a new dynamical system, the \emph{crease flow}, replacing the usual Hamiltonian flow. In particular,  we are able to describe the caustics that  arise during the evolution of the crease sets on null hypersurfaces, an issue that has not been properly addressed in the literature before. According to the crease evolutionary flow introduced here, the crease sets appear as its steady-state (i.e., fixed point) solutions, while their bifurcations describe the  possible types and topological transformations of the singularities very accurately.

To begin our analysis, for the system (\ref{sys1}) below we consider the transversal intersections of two null hypersurfaces $\mathcal{N}_i,i=1,2$, described by the equations $f_1=0,f_2=0$, respectively, with,
\begin{equation}\label{sys1}
\begin{split}
f_1&=-u+b_{AB}x^A x^B,\\
f_2&=-v+c_{AB}x^A x^B.
\end{split}
\end{equation}
We shall call these intersections `crease submanifolds', or just \emph{singularities}\footnote{no relation to geodesic incompleteness.}. Here $A,B=1,2$, while $t,x^1,x^2,z$ are Minkowski \emph{normal} coordinates,  $u,v$ are  null coordinates of the flat metric with $u=(t+z)/\sqrt{2},v=(t-z)/\sqrt{2}$, and $b,c$ denote quadratic polynomials, i.e., the second-order terms in the expansion in normal coordinates $x=x^1,y=x^2$ (the omission of the higher-order terms in (\ref{sys1}) will be further clarified below.) Also a time function will be denoted by $\tau$, and its level hypersurfaces $\Sigma_\tau$ will in general intersect the crease submanifolds transversally, or tangentially at pinch points.

In the previous works  (cf. e.g., \cite{gr23} and refs. therein),  such intersections are described by the geometry of the big wave front\footnote{in the context of Legendre maps, fronts are also called `equidistants'.} $\mathcal{W}=\mathcal{N}_1\cup\mathcal{N}_2$ and the hope is to provide a local description of the event horizon near an instant of merger, and its nucleation or collapse, or more generally their `metamorphoses'.  These phenomena correspond to, and can be described by, various bifurcation sets, singularities,  and their transfigurations at pinch points.

A basic feature of the analysis of these works is the use of Taylor series expansions corresponding to the truncation  (\ref{sys1})  to study non-caustic crease points where the big wavefront locally resembles the transverse intersection of two \emph{smooth} null hypersurfaces. Such expansions of course break down at caustic points and so cannot be used to describe the possible singularities of the null hypersurfaces, they are only valid in a small neighbourhood of a non-caustic crease point (cf. e.g., \cite{gr23}, Section 2). Indeed, at caustic points, the above argument based on the use of Taylor expansions breaks down and it is not clear how to make  physical sense and consistently describe  an evolution that develops such caustics.

This is exactly  where the present approach, based on bifurcation and singularity theory methods, enters. We provide a full and consistent answer for a complete description of the crease evolution that contains non-caustic \emph{as well as caustic} points. The results reported here offer a  global picture of the possible physical effects associated with intersections of null hypersurfaces including the formation and development of caustics. In particular, we deploy consideration and provide answers to the following  issues associated with this problem:
\begin{enumerate}
  \item The  crease structure and evolution problem  is at a most fundamental level a \emph{bifurcation problem}, especially  at those points where smoothness breaks down.
  \item To what extent do the low order terms in the Taylor expansion determine the qualitative nature of the problem regardless of higher-order terms  (i.e., `\emph{finite determinacy}' and the need to find the \emph{normal form} of the problem)?
  \item How can we determine the complete set of \emph{allowed perturbations} of the low order terms (\ref{sys1}) such that any perturbation of them becomes equivalent to some special form of a universal  family of perturbations (i.e., the construction of the \emph{versal unfolding})?
\end{enumerate}

In this work, we introduce the crease flow (in Eq. (\ref{sys2}) below) in order to  study the development of the caustics in the structure of the crease sets  and the associated continuous change of these singularities upon parameter variation. This approach is from  a bifurcation theory point of view using methods of the singularity theory of functions \cite{golu2,golu3,ar93}. In our sense, fronts will be defined more generally than in previous references as bifurcation sets corresponding to fixed points  of \emph{versal unfoldings} (i.e., versal families or universal deformations), cf. e.g., \cite{ar93}.

In the next Section, we introduce the new flow for the crease sets (replacing the usual Hamiltonian flow) and show that it can be formulated as a bifurcation problem with the radius of the intersection 2-spheres being the distinguished parameter. We also present a short summary of the main new results of this work. In Section 3, we discuss the nondegeneracy conditions associated with the crease submanifolds carried by the flow. In section 4, we write down the normal forms and versal unfoldings admissible for the proven reduction found in Sections 2, 3, while in Sections 5, 6, we find the allowed types singularities and built bifurcation diagrams of the creases which describe all possible metamorphoses of the various admissible singularities of the problem. Some further remarks are given in the last Section.

\section{The crease flow}
To introduce the crease flow, a new evolution system for studying the crease submanifolds,  it will prove more convenient to use the polar radius $r$ instead of the coordinate $z$ in the null coordinates, that is $u=t+r,v=t-r$, as it is done for example in Ref.  \cite{he}, pp. 118-9. This means that $u (v)$ represent incoming (outgoing) spherical waves traveling at the speed of light, while  intersections of a surface $u=\textrm{const.}$ with a surface $v=\textrm{const.}$ are 2-spheres.

However, this definition would spoil the polynomial character of the last terms in the right-hand-sides of (\ref{sys1}) by introducing trigonometric functions instead of the cartesian variables $x,y$ (this would also spoil the whole idea of using normal coordinates).
To remedy this, we further introduce  new \emph{synchronous} coordinates $T,X^1,X^2,X^3$ for the region $t>(x^2+y^2+z^2)^{1/2}$, which are defined as in \cite{pe72}, p. 53, namely, $T=(t^2-x^2-y^2-z^2)^{1/2},X^1=x/t,X^2=y/t,X^3=z/t$. Below we shall take $T$ to play the role of the time function $\tau$ above.

Then  we have,
$
T^2=uv-x^2-y^2,
$
and so on the spacelike hypersurfaces $T=\textrm{const}.$ (orthogonal to the timelike geodesics $X^1,X^2,X^3=\textrm{const}.$), the surfaces $u=\textrm{const}.,v=\textrm{const}.,$ lead to the equation $x^2+y^2=uv-T^2$, which are 2-spheres of radius $\lambda=\sqrt{uv-T^2}$. Similarly, for the other region $t<(x^2+y^2+z^2)^{1/2}$, $\lambda<0$, and the radius of the 2-spheres will be $-\lambda$.

This result implies that to a given point $(u,v)$ there corresponds a $\lambda$ value, and  for $\lambda_0\in\mathbb{R}$, the $\lambda=\lambda_0$-lines screen the $(u,v)$-plane. Therefore in the system (\ref{sys1}) we can use $\lambda$ as a distinguished parameter  in the place of the null coordinates $u,v$.

The \emph{crease flow} is then defined by the dynamical system,
\begin{equation}\label{sys2}
\begin{split}
\dot{x}&=\lambda+b_{AB}x^A x^B:\quad g_1 ,\\
\dot{y}&=\lambda+c_{AB}x^A x^B:\quad g_2,
\end{split}
\end{equation}
where the dot means differentiation with respect to the time function, and $x=x^1,y=x^2$ as above.

The significance of the crease  flow (\ref{sys2}) compared to the standard one (as the latter is described in the interesting works \cite{frie,stew91,kro,gr23} and refs. therein) for the description of the evolution of the crease sets, lies in the following new aspects of the present approach not to be found in  previous references:
\begin{enumerate}
  \item We provide a  description of the caustics formed during the evolution of the crease sets. This is done through a bifurcation  theory analysis of the emergence of the possible singularity types, rather than assuming their existence in the present problem.

  \item We do not use a hamiltonian (or contact) structure for the evolution of the null submanifolds. As a consequence,  we do not need to limit ourselves to the search of only Lagrangian or Legendrian singularities. In particular, we introduce and exploit newer results in singularity theory which where only fully formulated since the Legendrian approach to the characteristic initial value problem and related applications to the crease evolution first appeared. These results allow for an alternative interpretation of our findings using the Arnold-Kazaryan-Shcherbak theorem about the structure of singularities and bifurcation boundary sets.

  \item Hence, as a result, in our approach we  avoid  problems that arise in this respect (cf. e.g., \cite{gr23}, pp. 19-20), such as the non-invariance of the nullity of the wave front perturbations, or the impossibility of a consistent consideration of wave fronts in curved spacetimes (such approaches strictly hold only in an infinitesimal neighborhood of any spacetime point).  Instead,  we let the system (\ref{sys2}) itself point to the admissible types of singularities which may arise during evolution.

  \item We are able to consider the global bifurcations necessarily inherent in this problem. This is obviously not possible in an approach based on local flatness such as in \cite{gr23} and refs. therein.

  \item There are new singularities inherent in this problem which are necessarily suppressed in a Legendrian (or Lagrangian) approach. These  appear by necessity due to the structure of the crease sets which basically fixes the codimension (i.e., the number of parameters in the set of all stable perturbations) and so fully determines the problem. Hence in our approach there are no arbitrary elements.

  \item We  determine the precise ranges in parameter 3-space where different singularity types may appear and transfigure. This becomes possible only through the bifurcation theory approach adopted here.

  \item We do not assume that `the stability of metric perturbations is equivalent to that of the wave front', as it is done in previous works of this subject.
      In fact, such an assumption appears unattainable without further qualification (the system (\ref{sys1}), (\ref{sys2}) is totally unstable). Most importantly, we detail the exact notion of stability we use, namely, through the versal unfolding construction and the associated bifurcation diagrams (i.e., stability of the versal families).

  \item In distinction to all previous references, we provide explicit forms of the allowed normal forms of the problem. This is absolutely  necessary in order to avoid the use of Taylor series expansions (which necessarily break down at caustics), and also to predict  the singularities that are allowed in the present problem, rather than assume their forms.

  \item We provide the  transversality and nondegeneracy (i.e., dispersive) conditions necessary for a correct singularity classification, and for proving that the problem is reduced to a bifurcation problem thus admitting any singularities at all (i.e., implicit function theorem not holding).

  \item We describe precisely the possible metamorphoses of the caustics through their bifurcations in parameter and phase space. We note that in other works on this subject,  only two types of singularities are basically analysed and captured  (both are based on the $A_3$ singularities).  This is partly due to the restriction on a Legendrian approach. In our case, the situation is as we show below much more diverse and includes an augmented list of singularities and another augmented list of their metamorphoses.

  \item Our results on non-differentiability are of course related to the question of non-smoothness as was discussed in ref. \cite{kro}. However,  here the problem also contains a new aspect which is of a slightly different nature. This is so because due to the singular nature of the problem, we necessarily have to work with versal families, and these contain a great deal of non-differentiable behaviour. That is, although the variables are generally related  smoothly, it is not  possible to express the solutions as smooth functions of the parameters present in the unfoldings.
\end{enumerate}

\section{The dispersive conditions}
The significance of the system (\ref{sys2}) is that in the transverse phase space $(x,y)$,  a point  $(x_0,y_0)$ is an equilibrium solution of  (\ref{sys2}) iff there is  a crease submanifold, i.e., $g_1=g_2=0$ (or,  $f_1=f_2=0$ in (\ref{sys1})). This is a decisive step to formulating and proving stability of the crease evolution.
That is in some neighborhood of  such a point, the system  (\ref{sys1}) (or the vector field (\ref{sys2})) is a bifurcation problem in the two variables $x,y$, for the equation,
\be \label{g} g(x,y,\lambda)=(g_1,g_2)=0.
\ee
Thus one expects a qualitative change in the singularity submanifolds in this case, as the number of solutions  $n(\lambda)$ of Eq. (\ref{g}) changes as $\lambda$ varies. Here  $n(\lambda)$  denotes the number of pairs $(x,y)$ for which $(x,y,\lambda)$ is a solution of (\ref{g}) (i.e., an equilibrium solution of the crease flow (\ref{sys2})).

Of course, in general,  points on the crease submanifolds are moved by the geodesic flow. What is the advantage of using the crease flow (\ref{sys2}) as an alternative evolution equation to the geodesic flow in the study of the problem of the character of  singularities that may form during the evolution of  the crease submanifolds?

A main advantage is that the system (\ref{sys2}) is two-dimensional in the number $n$ of the state variables of the problem  (the variables are $x,y$, and  $n=2$). Since a bifurcation problem $g$ in $n$ state variables, has codimension at least $n^2-1$ (cf. \cite{golu2}), we find that the crease flow has codimension at least 3, instead of codimension at least 8 for the geodesic flow (where $n\geq 3$). Therefore studying the crease submanifolds and their singularities using the geodesic flow is equivalent to expecting these singularities only in a geodesic flow problem with at least eight parameters. Below we shall use the crease flow to study singularities and their bifurcations in the minimal codimension 3.

The system (\ref{sys2}) (equivalently Eq. (\ref{g}))  satisfies the following  conditions,
\beq
g(0,0,0) =0,\quad \frac{\partial g}{\partial (x,y)}=0, \quad &&\textrm{(non-hyperbolicity)}\label{non0}\\
\frac{\partial g}{\partial\lambda}\biggr\rvert_{(0,0)}\neq 0,  \quad && \textrm{(nondegeneracy-1)}\label{non1},
\eeq
where the second condition in (\ref{non0}) (that is the vanishing of the differential of $g$) means that the Jacobian matrix has two zero eigenvalues and is the zero matrix (no symmetry).

We note that these conditions for non-hyperbolicity, transversality, and  nondegeneracy (together with condition (\ref{non2}) below) are very important in the study of singularities because they are responsible for putting any such system in its possible normal forms, and also ensure the typicality of the solutions cf. \cite{golu2,golu3,ar83,gh83,wig,cot23}. They also imply, for instance,  that the structures will in general be non-smooth because, even thought the variables in (\ref{g}) are smooth and smoothly related, they guarantee that the variables $x,y$ cannot be expressed as smooth functions of the $\lambda$ when solving $g(x,y,\lambda)=0$, with $g$ given in (\ref{g}) (the function $g$  may in fact be infinitely differentiable).

We further assume that,
\be\label{non2}
\Delta\neq 0, \quad \textrm{(nondegeneracy-2)},
\ee
where $\Delta=B^2-4AC$, is the discriminant of the Jacobian $J_{b,c,0}$ of the system (\ref{sys2}) evaluated at $\lambda=0$. Here we have set,
\be\label{p,q}
b_{AB}x^A x^B=p_1x^2+p_2xy+p_3y^2,\quad c_{AB}x^A x^B=q_1x^2+q_2xy+q_3y^2,
\ee
so that
\be\label{quad}
J_{b,c,0}=Ax^2+Bxy+cy^2.
\ee
The coefficients $A, B, C$ as well as $\Delta$  can be routinely calculated as polynomials of the $p,q$'s. We may then set $h(x,y)=(p(x,y),q(x,y))$, with the quadratic polynomials $p,q$ as in Eq. (\ref{p,q}), so that we can write,  $g(x,y,\lambda)=h(x,y)+\lambda g_\lambda (0,0)+\dots$ (the dots indicate higher-order terms which vanish at the origin).

\section{The versal unfoldings}
We are now ready to apply singularity theory methods to the  system (\ref{sys2}) (or the equation (\ref{g})), since we may take this to mean what remains after center manifold (or, `Liapunov-Schmidt') reduction. This means that our results apply not only to the system (\ref{sys2}) but any equivalent system which can be written as a direct product of (\ref{sys2}) with a part that has no zero eigenvalue in its linearized Jacobian. This obviously includes, for instance, higher dimensional systems of dimension $(2+n)$, where $n$ is the dimension of a hyperbolic system in the product, such a hyperbolic subsystem plays no dynamical role after center manifold  reduction.

Accordingly, we may put (\ref{sys2}) in three inequivalent normal forms according to the sign  $\Delta$. Namely, when $\Delta<0$ and  (\ref{non1}) holds, the system (\ref{sys2}) is equivalent to,
\begin{equation}\label{nf1}
\begin{split}
\dot{x}&=x^2-y^2+\lambda,\\
\dot{y}&=2xy,
\end{split}
\end{equation}
while when $\Delta>0,$ the system (\ref{sys2}) becomes topologically equivalent to the following normal forms,
\begin{equation}\label{nf2}
\begin{split}
\dot{x}&=x^2-\lambda,\\
\dot{y}&=y^2-\lambda,
\end{split}
\end{equation}
and,
\begin{equation}\label{nf3}
\begin{split}
\dot{x}&=x^2+\lambda,\\
\dot{y}&=y^2-\lambda,
\end{split}
\end{equation}
under the additional condition that $g_\lambda (0,0)\times \eta_i\neq 0$, that is $g_\lambda (0,0)$ is not parallel to the two directions $\eta_i,i=1,2$, where the quadratic form (\ref{quad}) vanishes (in this case, there are coordinates such that $p=x^2,q=y^2$).

These normal forms lead to the following versal unfoldings respectively (see, e.g., \cite{golu2}),
\begin{equation}\label{va}
\begin{split}
\dot{x}&=x^2-y^2+\lambda +2\mu_1 x-2\mu_2 y,\quad\dot{y}=2xy+\mu_3, \\
\dot{x}&=x^2-\lambda +2\mu_1 y -\mu_3,\quad\dot{y}=y^2-\lambda +2\mu_2 x+\mu_3, \\
\dot{x}&=x^2+\lambda +2\mu_1 y +\mu_3,\quad\dot{y}=y^2-\lambda +2\mu_2 x+\mu_3,
\end{split}
\end{equation}
where the $\lambda$ is the distinguished bifurcation parameter of the problem, and the $\mu_i$'s are the unfolding parameters.

We note that all three versal families are 3-parameter ones, meaning that the problem is now reduced to being one of codimension 3 in all cases. This result decisively affects the bifurcation diagrams to be constructed below.

\section{Types of singularities}
Below, we shall discuss implications of the crucial fact that all three versal unfoldings are of codimension 3, that is they contain three unfolding parameters.

The first conclusion is that the complexity of the singularity manifolds and their bifurcations is limited to  those corresponding to either 1-, or 2-, or at most 3-parameter cases. This means, in particular,  that the singularities that can generally appear in the present problem can only be of the following stable kinds (i.e., nearby caustics have the same form, cf. eg., \cite{ar83}, pp. 248-50):
\begin{enumerate}\label{enu}
  \item Folds (1-parameter families)
  \item Cusps and self-intersections (2-parameter families)
  \item Swallowtails, elliptic (pyramid) and hyperbolic (purse) umbilics (3-parameter families).
\end{enumerate}
(These also correspond to the standard classification of the transition varieties, that is the bifurcation boundaries of the strata in the bifurcation diagrams, and can be described as in pp. 409-11  of  \cite{golu2}, but we do not follow the analysis of that reference here.) All other objects with hyperbolic (i.e., non-dispersive) singularities are off these submanifolds. In addition, any singularity of codimension higher than 3 (in our problem of minimal codimension 3) can be eliminated and resolved into the standard ones enumerated above.

However, here and in the next Section we shall follow the approach of Ref. \cite{cot23}, where the unions of transversally intersecting null hypersurfaces occur in the bifurcation diagrams of the generic versal families (\ref{va}). This has the added advantage that the singularities appear in the bifurcation \emph{boundaries} of these diagrams, we shall study an example of this in the next Section.

However, in the present case of 3-parameter families, such singularities have forms governed by the Arnold-Kazaryan-Shcherbak theorem \cite{ar93} which gives possible different descriptions of them. In particular, in the $\mu$-parameter 3-space (i.e., $\mu=(\mu_1,\mu_2,\mu_3$)) the  boundary singularities will be described by polynomials  with multiple roots as follows:
\begin{enumerate}
\item a swallowtail, (`$A_3$' singularity) $s^4+\mu_1 s^2 +\mu_2 s+\mu_3$,
\item a folded umbrella (`$D_4$' singularity), $s^4+\mu_1 s^3 +\mu_2 s^2+\mu_3$
\item a pair of Whitney umbrellas, $s^4+3(\mu_1+\mu_2) s^2 + \mu_3 s -3\mu_1\mu_2.$
\end{enumerate}
This completes the structure of the bifurcation sets for our problem. We refer the reader to Figs. 90-97 of \cite{ar93} for pictures of the various singularities that are allowed in the present problem as we detailed above.

Of course, we are eventually interested in the \emph{bifurcations} that possibly occur when crossing a bifurcation boundary (given by the singularities above). We shall describe an example of this in the next Section.

\section{The bifurcation diagram in a simple example}
The system (\ref{sys1})  moves around in parameter 3-space according to the bifurcation diagrams corresponding to the versal unfoldings (\ref{va}). These are principally governed by bifurcations which occur at the crossings with the bifurcation sets, and depend on the signs of the unfolding parameters in the parameter space regions. The bifurcations involved presently are saddle-nodes, pitchforks,  and Hopf bifurcations.

Let us consider, as an example, the bifurcation problem  of the versal unfolding,
\be \label{red1}
\dot{x}=\lambda +2\mu_1 x +x^2-y^2,\quad \dot{y}=2xy,
\ee
obtained by setting $\mu_2=\mu_3=0$ in the bifurcation problem (\ref{va}a). This may be viewed as a projection of the problem (\ref{va}a) to the  $\mu_1$-axis of the 3-parameter space, while the other projections, where $\mu_1=\mu_2=0$, and $\mu_1=\mu_3=0$, are treated similarly, and together they completely solve the full problem (\ref{va}a). Problems (\ref{va}b,c) can be treated with similar methods.

The system (\ref{red1}) itself  is a 2-parameter versal unfolding of the totally degenerate system $(x^2-y^2, 2xy)$ obtained  by setting $\lambda=\mu_1=0$ (for an alternative study of this system in terms of singularity theory, we refer to \cite{golu2}, Chap. IX).  The  approach  followed here is however, to apply  bifurcation dynamics, and via  a parameter-coordinate shift $X=x-x_0, Y=y-y_0$, and suitable rescalings of the parameters $(\lambda,\mu_1)\to (\alpha,\beta)$, with $\alpha=\lambda-x_0^2,\beta=2x_0, \mu_1=-x_0$, it is straightforward to show that (\ref{red1}) is equivalent to the versal unfolding,
\begin{equation}\label{vaB}
\begin{split}
\dot{X}&=\alpha+X^2-Y^2,\\
\dot{Y}&=\beta Y+2XY.
\end{split}
\end{equation}

This system (in different variables) was studied in \cite{cot23}, Sect. 6, where it was shown that it  leads to the bifurcation diagram given in Fig. 11 in that reference (see also \cite{gh83,wig} for other equivalent forms of this versal unfolding). The parameter diagram is as in Figure 9 of \cite{cot23} with axes $\alpha, \beta$ (and one sets $n=2$ in that reference). The bifurcation boundaries are the positive and negative $\beta$-axis, the parabola $\alpha=-\beta^2/4$, and the $\nu$-line in the fourth quadrant (cf. \cite{cot23}).

In particular, we find the following bifurcations which occur when crossing one of the bifurcation boundaries:
\begin{enumerate}
\item a pair of saddle-node bifurcations when crossing any of the two ($\alpha=0$)-semiaxes,
\item a pair of pitchfork bifurcations when crossing either branch of the parabola $\alpha=-\beta^2/4$, and
\item a Hopf bifurcation to a stable limit cycle when crossing the positive $\beta=0$-semiaxis.
\end{enumerate}

These bifurcations move the system globally in parameter space, and generally describe \emph{defocusing} solutions in the metamorphoses of the phase portraits of the system. In particular,  to each parameter point in the $(\alpha, \beta)$ space, there corresponds a phase portrait in the $(X,Y)$
variables as studied in Ref. \cite{cot23}, Section 6. As the parameter point moves in the $(\alpha, \beta)$ plane, the phase portrait is smoothly transfigured continuously to new forms which describe the overall possible metamorphoses of the  crease sets.

We note that the swallowtail and umbrella singularities given above in the previous Section appear when crossing the bifurcation sets in the bifurcation diagram, and describe the continuous changes of the system from one state to another through the transformation of the phase portraits. These are in turn dictated by the crease evolution equations as described above.

In the `projected' versal unfolding (\ref{vaB}), one sees the evolution of the projections of the singularities when $\mu_2=\mu_3=0$ (the whole image will be a `superposition' of all three projections together to generate the full problem (\ref{va}a)).

A particularly interesting aspect of the ensuing  metamorphoses is the bifurcation of the crease into a stable limit cycle. This is an important feature of the bifurcating diagrams, namely, the existence of a Hopf bifurcation as the parameter point $(\lambda,\mu_1)$ varies suitably. It describes a phenomenon in which the crease (a steady state solution of the evolution equations describing the crease flow) evolves into a periodic orbit - we have given sufficient conditions for the occurrence of this stable configuration.

The resulting scenario leads to an evolving,  in principle observable, pattern  of bifurcating caustics corresponding to intersections of null hypersurfaces in spacetime. This is increasingly difficult to describe (let alone picture!), for at each small spacetime region one observes the joined effects of the three bifurcations mentioned above (saddle-node, pitchfork, and Hopf)  acting on the following caustics: swallowtail, folded Whitney, and a pair of `embedded in each other' (i.e., combined)  Whitney singularities, all three being steady states of the crease flow (\ref{sys2}).

Based on the  standard physical interpretation of the bifurcations, any such steady state solution:
\begin{enumerate}
\item can be created (or destroyed) in a saddle-node bifurcation (recall here the important role played by the `ghost effect' - the square-root scaling law for the time to its formation of destruction, cf. \cite{cot23}),
\item may blow up or decay (undergoing a 1st- or 2nd-order phase transition) according to a pitchfork, or,
\item may become a closed orbit in a Hopf bifurcation.
\end{enumerate}
As an illustration, let us suppose that we have a solution \emph{created} in a saddle-node (from `nothing', i.e., from a previous state of no equilibria, cf. \cite{cot23}), that is we choose a particular direction in parameter space for (one of) the saddle-nodes. It may change its size in a pitchfork, or loose its stability and eventually become a closed orbit in a Hopf bifurcation. As the parameter changes differently at different spacetime points, these effects occur simultaneously, thus offering a very rich picture for the evolution of the crease sets. Since in the crease flow the $x,y$ are two of the spacetime coordinates, one may think of the evolution as describing  \emph{2-sections} of  deformed `3-structures' of  a swallowtail or one of the Whitney-like singularities (for instance, using the $z$ as the third coordinate, the latter are given by the forms $x^2=zy^2$ (standard Whitney), and $x^2=z^3y^2$ (folded)), and  evolving in proper time through the bifurcations.

One thus envisages a kind of `fluid'-like picture of  such  structures smoothly evolving together and continuously changing their  forms, all such transfigurations being smooth. We note that once a stable limit cycle is formed, it can never be destroyed but may only become `invisible' (as was detailed  in \cite{cot23}, also cf.  refs. therein). The overall picture of the resulting evolution of the caustics is structurally stable, as this follows from the stability of any of the three versal families.

\section{Conclusion}
In this Letter, we have studied the crease evolution on null hypersurfaces by introducing a novel flow and  its bifurcation and singularity theory  properties. The crease flow leads to a number of  effects, most notably the description of the caustics formed in the evolution.   It reformulates the evolution problem for the creases as a  bifurcation problem, and this has a number of advantages (compared to the usual treatment through an \emph{ad hoc} hamiltonian system) not met in previous literature on this subject before.

The basic nature of the evolution of creases is related to the problem being a degenerate one in the sense of bifurcation theory. This means that certain conditions have to be fulfilled for the evolution of the crease system to be consistent. We have in particular derived  dispersive, non-degeneracy,  and transversality conditions for this purpose. These in turn lead to the normal forms of the system and to the versal families that fully describe the evolution of the system during caustic formation and development.

The crease evolution is then characterized by bifurcation boundaries and other strata in the obtained bifurcation diagrams of the problem. The nature of the allowed singularities is determined by number of parameters present in the versal unfolding (three in this case), and correspond to swallowtails and specific types of Whitney umbrellas. These  then evolve according to the allowed bifurcations of the problem.

Our results eventually describe the behaviour of the system (\ref{vaB}) along the corresponding axis in the parameter space of each versal unfolding of  the problem (\ref{red1}) as documented in the bifurcation diagrams, and therefore  highlight the evolution of the \emph{projections} of the singularities discussed earlier as well as their metamorphoses obtained every time the system bifurcates upon parameter variation (similarly for the other two versal families (\ref{va}b,c)).

One further expects that in situations of axisymmetry (i.e., equivariance with respect to some symmetry like $x\to \pm x,y\to\pm y$), the codimension will actually drop below 3 so that the problem  becomes somewhat simpler to handle (but more varied).

Because of the possibility of saddle connections in the bifurcation diagrams of the previous Section (cf. \cite{cot23}), one expects them to be near the caustics of type $D_4$ in the classifications of catastrophe theory (Thom's program, \cite{ar94}). However, these singularities typically depend on 4 parameters instead of 3 that we have in this paper, and so their typicality in the present problem is an open issue. In our 2-parameter problems (the `projections' in the previous Section), Thom's program is of course also realized but the deeper relations between the metamorphoses of the phase portraits and those of the catastrophes is still elusive for the crease problem.

As this work describes  properties of the crease flow (and its versal unfoldings) and its singularities on null hypersurfaces, it  may also offer a glimpse to the particular problem of the  structure and evolution of black hole horizons. We note here that the achronality of some of the possible creases proved in previous references is not really helpful in this respect,  because it only relates to a `static' aspect of this problem. We need to know exactly how achronality is respected by the crease flow during evolution as the system bifurcates through the caustics, not only instantaneously.

We lastly note that the present approach to crease evolution is potentially also a new method to the characteristic initial value problem in general relativity (in a similar way as the Hamiltonian system approach of \cite{frie} were). This is of course not analyzed here at all, but provides a further possibility into approaching certain aspects of the structure of general relativity in a novel way.

A more detailed discussion of these and related results will be given elsewhere.

\section*{Acknowledgments}
We thank an anonymous referee for constructive comments. The author thanks Gary Gibbons for discussions, and Harvey Reall and Max Gadioux  for many useful discussions about their work.
A Visiting Fellowship to Clare Hall, University of Cambridge, is gratefully  acknowledged, and the author further thanks Clare Hall for  its warm hospitality and partial financial support.
This research  was funded by RUDN University,  scientific project number FSSF-2023-0003.

\end{document}